\documentclass{article}

\PassOptionsToPackage{numbers, compress}{natbib}

\usepackage[final]{neurips_2023}

\usepackage[utf8]{inputenc} %
\usepackage[T1]{fontenc}    %
\usepackage{hyperref}       %
\usepackage{url}            %
\usepackage{booktabs}       %
\usepackage{amsfonts}       %
\usepackage{nicefrac}       %
\usepackage{microtype}      %
\usepackage{xcolor}         %
\usepackage{multirow}       %
\usepackage{adjustbox}      %
\usepackage{amsmath}        %

\usepackage[textsize=tiny]{todonotes}
\usepackage[nameinlink]{cleveref}
\usepackage{bm}

\title{Unsupervised Musical Object Discovery from Audio}

\author{
   Joonsu Gha$^{1}$ ~ Vincent Herrmann$^{1}$ ~ Benjamin Grewe $^{2}$ \\ 
   \textbf{J\"urgen Schmidhuber $^{1,3}$} ~ \textbf{Anand Gopalakrishnan$^{1}$} \\
  $^1$The Swiss AI Lab, IDSIA, USI \& SUPSI, Lugano, Switzerland \\
  $^2$ Institute of Neuroinformatics, ETH Zurich, Z\"urich, Switzerland \\
  $^3$AI Initiative, KAUST, Thuwal, Saudi Arabia \\
   \texttt{joonsu.gha@usi.ch} \\
   \texttt{bgrewe@ethz.ch} \\
   \texttt{\{vincent.herrmann, juergen, anand\}@idsia.ch}
}

\begin{document}

\maketitle

\begin{abstract}
Current object-centric learning models such as the popular SlotAttention architecture allow for unsupervised visual scene decomposition. Our novel \emph{MusicSlots} method adapts SlotAttention to the audio domain, to achieve unsupervised music decomposition. Since concepts of opacity and occlusion in vision have no auditory analogues, the softmax normalization of alpha masks in the decoders of visual object-centric models is not well-suited for decomposing audio objects. \emph{MusicSlots} overcomes this problem. We introduce a spectrogram-based multi-object music dataset tailored to evaluate object-centric learning on western tonal music. \emph{MusicSlots} achieves good performance on unsupervised note discovery and outperforms several established baselines on supervised note property prediction tasks.\footnote{Official code repository: \url{https://github.com/arahosu/MusicSlots}}
\end{abstract}

\section{Introduction}
\label{sec: intro}
Human infants learn to group the feature cues of incoming visual stimuli into a set of meaningful entities \citep{spelke1990object}.
This capacity to integrate feature cues into a ``unified whole'' \citep{koffka1935principles, kohler1967gestalt} extends beyond visual perception to the auditory domain \citep{kubovy2001auditory, griffiths2004whatauditoryobject, bizley2013whatwherehow}. 
The notion of `object files' \citep{kahneman1992reviewing} that capture this visual feature integration has been posited to extend to the auditory domain \citep{hall2000evidence, zmigrod2009auditory} as well.
Recently, there has been growing interest in developing unsupervised deep learning models for perceptual grouping ( ``object-centric learning'') in the visual domain \citep{eslami2016attend, greff2016tagger, greff2017neural, steenkiste2018relational, kosiorek2018sequential, stanic2019rsqair, crawford2019spair, burgess2019monet, greff19aiodine, locatello2020slotattn, engelcke2020genesis, lin2020space, singh2022illiterate, kipf2022savi}. 
Such object-centric models bias the underlying structure of machine perception to be human-like by modeling the scene as a composition of objects.
However, the object-centric learning literature has primarily focused on perceptual grouping task for vision (images/video) and extensions to the auditory domain remain largely unexplored.
Therefore, we focus on extending object-centric models for the auditory, specifically musical domain. 
To the best of our knowledge, no prior work has applied unsupervised object-centric models to the problem of unsupervised music decomposition.

\begin{figure*}[t]
\centering
\includegraphics[width=0.74\textwidth]{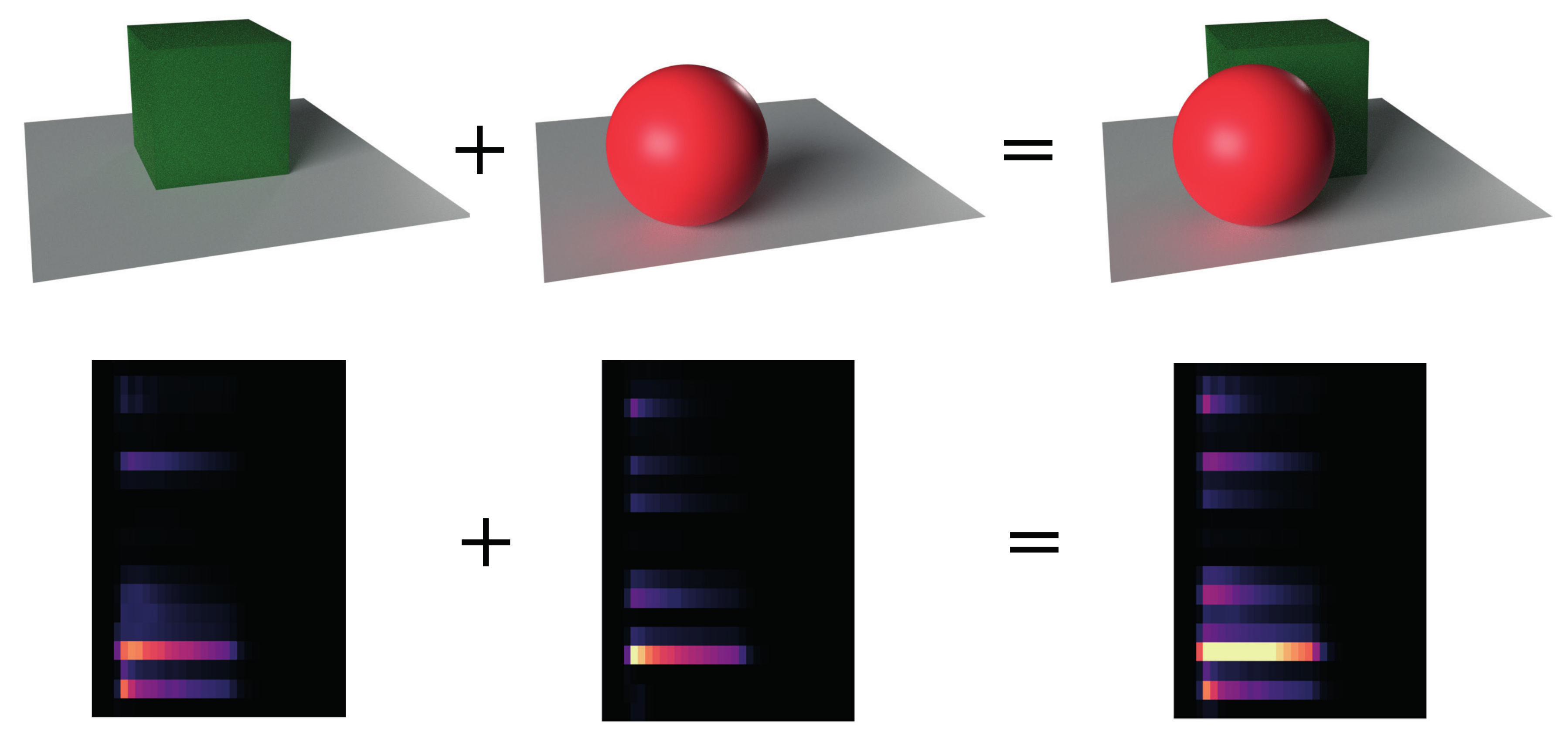}
\caption{Illustration of the effects of opacity and occlusion in visual and auditory (spectral) domains.}
\label{fig: opacity-occlusion-illustration}
\end{figure*}

Western tonal music serves as a suitable form of auditory signal for our study as its building blocks are symbolic units such as notes, chords or phrases, which themselves are organized into more complex structures using rich grammars \citep{lerdahl1983generative}. 
We investigate if object-centric models (SlotAttention \citep{locatello2020slotattn}), highly successful in visual grouping, are able to segregate constituent units of a musical score given its spectrogram in a fully unsupervised manner. 
The auditory modality poses unique challenges as the underlying structure of auditory objects differs from their visual counterparts.
For instance, the concepts of occlusion and opacity in the visual domain have no auditory analogues. 
If two auditory objects occupy some overlapping spectral regions, their composition would approximately result in the additive combination of their power spectra in these regions.
In contrast, on the visual domain where two opaque objects cannot occupy the same spatial location and one will necessarily occlude the other e.g. the red ball in front of green cube in \Cref{fig: opacity-occlusion-illustration}.
Therefore, in the visual case every pixel is naturally assumed to belong to only one object while for audio this assumption does not hold true. 

We propose \emph{MusicSlots}, an autoencoder model to decompose a chord spectrogram into its constituent note spectrograms (objects) in a fully unsupervised manner. 
We show that \texttt{Softmax} normalization (across slots) of alpha masks in the SlotAttention decoder \citep{locatello2020slotattn} is not well-suited for discovering musical objects as it assumes that the feature at any spatial location belongs to only one object (slot) which is invalid for the audio domain.
Further, we introduce a multi-object music dataset tailored to evaluate object-centric learning methods for music analogous to its visual counterpart \citep{kabra2019multiobjectdata}. 
Our dataset consists of chord spectrograms (taken from Bach-Chorales \citep{boulanger2012jsbdata} and Jazznet \citep{adegbija2023jazznet}), constituent note spectrograms and corresponding ground-truth binary masks. 
Finally, we show that our \emph{MusicSlots} model achieves good performance on unsupervised note discovery,
outperforms several baseline models (VAE, $\beta$-VAE, AutoEncoder, supervised CNN) on supervised note property prediction task and shows generalizes better to unseen note combinations and number of notes.

\section{Method}
\label{sec: method}
\begin{figure*}[t]
\centering
\includegraphics[width=0.95\textwidth]{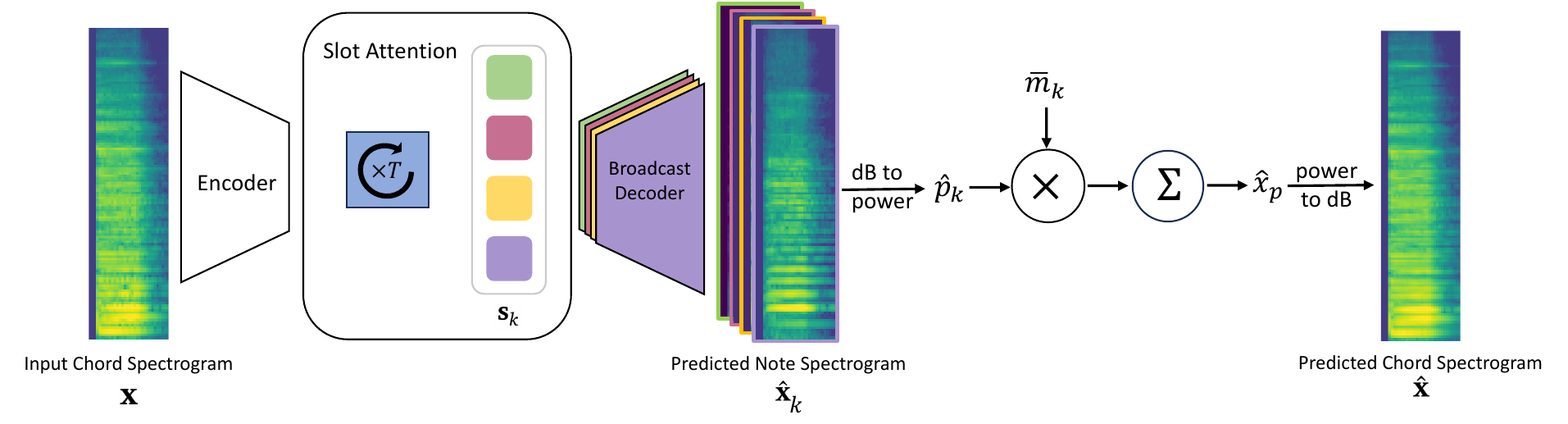}
\caption{MusicSlots model consists of 3 modules --- Encoder, SlotAttention and Broadcast Decoder.}
\label{fig: model-diagram}
\end{figure*}

Given a mel-scale spectrogram representation $\mathbf{x} \in \mathbb{R}^{D_{in} \times h \times w}$ of a musical chord, our goal is to decompose it into its constituent note-level spectrograms $\mathbf{x}_k \in \mathbb{R}^{D_{in} \times h \times w} \quad \forall k=\{1, 2,..., K\}$ and learn their associated representations (slots) $\mathbf{s} \in \mathbb{R}^{{D}_{s} \times K}$. 
Here, $D_{in}$ denotes the number of channels in the spectrogram, $D_{s}$ the slot size, $h, w$ the number of frequency bins and time window of the input spectrogram respectively and $K$ the total number of slots. 
Our proposed \emph{MusicSlots} model is an autoencoder which consists of three modules (see \Cref{fig: model-diagram}). 
An encoder module (CNN) to extract features from the input spectrogram, slot attention module to group input features to slots and decoder module to reconstruct the chord spectrogram from the slot representations.

\paragraph{Encoder.}
The encoder module consists of a CNN backbone to extract features $\mathbf{h} \in \mathbb{R}^{{D}_{f} \times h \times w}$ from the input chord spectrogram $\mathbf{x}$. 
Learnable positional embeddings $\mathbf{p} \in \mathbb{R}^{D_f \times h \times w}$ (see \Cref{sec: model-details} for more details) are added to the output features $\mathbf{h}$ from the final convolutional layer. 

\paragraph{Slot Attention.} 
The slot attention module learns to map a set of $ N = h \cdot w$ input features onto a set of $K$ slots using an iterative attention mechanism (Algorithm 1 from \citet{locatello2020slotattn}) implemented via key-value attention \citep{vaswani2017attention} and recurrent update function.
The input features $\mathbf{h}$ are projected to a set of keys $\mathbf{k}$ and values $\mathbf{v} \in \mathbb{R}^{D_s \times N}$ using separate linear layers. 
Each slot $s_k$ is initialized as independent samples from a Gaussian distribution $\mathcal{N}(\mu_k, \sigma_k)$ with separate learnable mean $\mu_k \in \mathbb{R}^{D_s}$ and standard-deviation $\sigma_k \in \mathbb{R}^{D_s}$. 
Then at each iteration $t = \{1, ..., T\}$, slots compete to represent elements of the set of features using standard key-value attention (with features as keys \& values and slots as queries) except with softmax normalization applied across the slots.   
We use a recurrent network (specifically a GRU \citep{cho2014gru, gers2000lstmforget}) and residual MLP \citep{locatello2020slotattn}) to update the slots with weighted values as inputs and slots at $t-1$ as hidden states of the RNN.  
Further, our \emph{MusicSlots} model also adopts recent improvements to SlotAttention such as implicit differentiation \citep{chang2022fixedpoints}. 
\paragraph{Decoder.}
Each slot $s_k$ is decoded independently by the spatial broadcast decoder \citep{watters2019spatial} (see \Cref{fig: model-diagram}) using several de-convolutional layers. 
First, slots are broadcasted onto a 2D grid (independently) and learnable positional embeddings are added.
The decoder outputs the reconstructed note-level spectrogram $\mathbf{\hat{x}}_k \in \mathbb{R}^{D_{in} \times h \times w}$ and un-normalized (logits) alpha mask $\mathbf{m}_k \in \mathbb{R}^{1 \times h \times w}$. 
The individual slot-wise spectrograms and normalized masks $\mathbf{\overline{m}}_k = f_{norm} (\mathbf{m}_k) \in \mathbb{R}^{1 \times h \times w}$ are alpha composited to give the predicted power-scale chord spectrogram $\mathbf{\hat{x}}_p = \sum_{k=1}^{K} \mathbf{\hat{p}}_k \odot \mathbf{\overline{m}}_k$ where $\mathbf{\hat{p}}_k$ is the power-scale note spectrogram, $\odot$ is an element-wise multiplication and $f_{norm}$ is the normalization function.
This composition operation needs to carried out in power-scale followed by conversion back to decibel scale to get the predicted chord spectrogram $\mathbf{\hat{x}}$ as follows:
\begin{equation}
    \mathbf{\hat{p}}_k = 10^{\mathbf{\hat{x}}_k / 10} \quad ; \quad \mathbf{\hat{x}}_p = \sum_{k=1}^{K} \mathbf{\hat{p}}_k \odot \mathbf{\overline{m}}_k \quad ; \quad \mathbf{\hat{x}} = 10 \log_{10} \Big (\frac{\mathbf{\hat{x}}_p}{\mathbf{\hat{x}}_{p0}} \Big) \textrm{dB}
    \label{eqn:postprocessing}
\end{equation}
where $\mathbf{\hat{p}}_k$ is the power-scale note spectrogram, $\mathbf{x}_p$ is the chord spectrogram in power scale and $\mathbf{\hat{x}}_{p0}$ is the reference power. 
Further, a crucial modification required to adapt SlotAttention to the audio domain, is the choice of this normalization function $f_{norm}$ for alpha masks. 
For auditory objects, as illustrated in \Cref{fig: opacity-occlusion-illustration} notions of occlusion and opacity are invalid which means that its feasible for one or more notes (slots) to contribute to the power at a spatial location (frequency bin and time) in the spectrogram.
Contrarily, in the visual domain its necessarily the case that every pixel belongs to only one object. 
Therefore, we experiment with alternatives such as \texttt{Sigmoid} and not using any alpha masks in the broadcast decoder.
We train our \emph{MusicSlots} model using the MSE between the predicted and input chord spectrogram $\mathcal{L} = ||\mathbf{x} - \mathbf{\hat{x}}||_2^2$.
For more details on model architecture and training hyperparameters please refer to \Cref{sec: model-details} and \Cref{sec: training-details} respectively. 

\section{Related Work}
\label{sec: related-work}
Learning music representations from audio has been explored using self-supervised techniques such as autoencoders \cite{caillon2022rave, zeghidour2021soundstream, li2022mapmusic2vec} and contrastive methods \cite{spijkervet2021contrastive, mccallum2022supervised, choi2022proper, wang2022learning} or weak supervision from different modalities \cite{weston2011multi, park2017representation, manco2022weak, chen2022text}. %
Related to our work, `Audioslots' \citep{reddy2023audioslots} applies object-centric models to the audio domain for blind source separation.
However, their model is strongly supervised as it is trained using MSE loss between predicted and matched ground-truth individual source spectrograms.
Others have applied slot-based object-centric models beyond vision to learn modular action sequences for RL agents \citep{gopalakrishnan2023slottar} or robotic control policies \citep{zhou2023modularrobot}. 
\looseness=-1
\section{Results}
\label{sec: results}
We describe details of our multi-object music dataset followed by results on unsupervised note discovery and supervised note property prediction tasks. 
\paragraph{Multi-Object Music Datasets.}
To evaluate the efficacy of our proposed \emph{MusicSlots} model on the unsupervised music decomposition task, we need a dataset of musical scores in the spectral format and its decomposition into sub-parts i.e. part-level spectrograms and binary masks (see \Cref{fig: binary mask visualization}). 
We introduce synthetic multi-object datasets to specifically evaluate object-centric learning methods on the music domain analogous to its visual counterpart \citep{kabra2019multiobjectdata}.
First, we extract chords (MIDI tokens) from Bach-Chorales (JSB) \citep{boulanger2012jsbdata} and JazzNet \citep{adegbija2023jazznet} datasets.
Next, we synthesize the audio waveform and its spectrogram for these chords (see \Cref{sec: multi-object music dataset details} for more details).  
Our dataset consists of two variants --- i) single-instrument: all notes in a chord played by the same instrument, ii) multi-instrument: different notes in a chord played by different instruments.
We create out-of-distribution test splits that measure generalization to unseen note combinations and number of notes in a chord. 
The test split in Bach Chorales contains chords with known notes in unknown combinations (w.r.t train/validation splits) whereas in JazzNet it contains only chords with four notes, while training and validation splits consist of two and three-note chords (see \Cref{sec: multi-object music dataset details} for more details). 

\paragraph{Note Discovery.}
We train three variants of our \emph{MusicSlots} model with different choices for $f_{norm}$ --- i) \texttt{Softmax} (MusicSlots-soft) ii) \texttt{Sigmoid} (MusicSlots-sigm) iii) no alpha mask usage (MusicSlots-none). 
We refer to \Cref{table: note discovery training} and \Cref{sec: model-details} for model/training details. 
We quantitatively measure note (object) discovery performance of our models using the best matched note-level MSE of spectrograms and mean IoU scores of binary masks.
\Cref{table: note discovery results} shows the note discovery results of \emph{MusicSlots} on multi-instrument versions of JSB and JazzNet datasets. 
We see that the \emph{MusicSlots} without any alpha masking is competitive with \texttt{Sigmoid} normalization and these alternatives show significant performance gains over the default \texttt{Softmax}. 
Further, we observe that training on multi-instrument chord datasets is beneficial for better decomposition quality (compare with single-instrument in \Cref{sec: add-results}).
We show samples of good decomposition and some failure cases of our \emph{MusicSlots} model in \Cref{sec: add-viz}.
\looseness=-1

\begin{table}[h]
\caption{Note discovery results on multi-instrument BachChorales (JSB) and JazzNet datasets for MusicSlots models. Mean and std-dev. are reported across 5 seeds.}
\label{table: note discovery results}
\begin{center}
\begin{adjustbox} {width=0.62\textwidth}
\begin{tabular}{c|c|c|c}
\toprule
 Datasets & Mask Norm. & Note MSE $\downarrow$ & mIoU $\uparrow$ \\ \midrule
\multirow{2}{*}{JSB-multi} & MusicSlots-soft & 59.34 ${\scriptstyle \pm 22.01}$ & 0.79 ${\scriptstyle \pm 0.04}$ \\
 & MusicSlots-sigm & \textbf{13.07} ${\scriptstyle \pm 0.80}$ & 0.90 ${\scriptstyle \pm 0.01}$ \\
 & MusicSlots-none & 13.47 ${\scriptstyle \pm 0.90}$ & \textbf{0.91} ${\scriptstyle \pm 0.01}$ \\ \midrule
 \multirow{2}{*}{JazzNet-multi} & MusicSlots-soft & 33.58 ${\scriptstyle \pm 2.08}$ & 0.84 ${\scriptstyle \pm 0.01}$ \\
 & MusicSlots-sigm & \textbf{18.53} ${\scriptstyle \pm 0.83}$ & \textbf{0.91} ${\scriptstyle \pm 0.01}$ \\ & MusicSlots-none & 19.95 ${\scriptstyle \pm 1.89}$ & 0.90 ${\scriptstyle \pm 0.01}$ \\
 \bottomrule
\end{tabular}
\end{adjustbox}
\end{center}
\end{table}

\paragraph{Note Property Prediction.} 
We train a linear classifier with cross-entropy loss (see \Cref{sec: model-details} for details) to predict the properties (MIDI pitch value and instrument type) of all notes in a chord from the frozen (pre-trained) latent representations.  
We use the classification accuracy as the evaluation metric for this task wherein a chord is considered to be correctly classified if and only if all its note pitch values and instrument identities are correctly predicted.
The supervised CNN baseline uses the same encoder module as \emph{MusicSlots} followed by a two layer MLP and trained in a supervised manner to predict the note properties given the chord spectrogram.  
\Cref{table: note property prediction results} shows the note property results for our \emph{MusicSlots} model against various baseline models. 
We see that our \emph{MusicSlots} model outperforms several unsupervised baseline models (AutoEncoder/VAE/$\beta$-VAE).
Surprisingly it also outperforms the supervised CNN baseline which is explicitly trained end-to-end to solve the task.  
Further, \emph{MusicSlots} shows a greater degree of generalization to unseen note combinations on the test splits of JSB-multi and different number of notes in a chord on JazzNet-multi respectively.
\looseness=-1

\begin{table}[h]
\caption{Note property prediction performance of \emph{MusicSlots} compared to Autoencoder, VAE, $\beta$-VAE and supervised CNN baseline models. Mean and std-dev. are reported across 5 seeds.}
\label{table: note property prediction results}
\begin{center}
\begin{adjustbox}{width=0.71\textwidth}
\begin{tabular}{c|cc|cc}
\toprule
\multirow{2}{*}{Models} & \multicolumn{2}{c|}{JSB-multi} & \multicolumn{2}{c}{JazzNet-multi} \\ \cline{2-5}
 & Val-Acc. & Test-Acc. & Val-Acc. & Test-Acc. \\
\midrule
Supervised CNN & 93.49 ${\scriptstyle \pm 2.10}$ & 93.03 ${\scriptstyle \pm 2.10}$ & 96.47 ${\scriptstyle \pm 0.79}$ & 71.22 ${\scriptstyle \pm 2.93}$ \\
AutoEncoder & 95.02 ${\scriptstyle \pm 0.09}$ & 93.62 ${\scriptstyle \pm 0.39}$ & 92.91 ${\scriptstyle \pm 0.26}$ & 71.37 ${\scriptstyle \pm 1.07}$ \\
VAE & 96.66 ${\scriptstyle \pm 0.34}$ & 96.21 ${\scriptstyle \pm 0.35}$ & 94.37 ${\scriptstyle \pm 0.55}$ & 71.55 ${\scriptstyle \pm 4.77}$ \\
$\beta$-VAE & 97.85 ${\scriptstyle \pm 0.13}$ & 97.42 ${\scriptstyle \pm 0.06}$ & 97.00 ${\scriptstyle \pm 0.37}$ & 81.53 ${\scriptstyle \pm 0.77}$ \\
MusicSlots-none & \textbf{98.13} ${\scriptstyle \pm 0.16}$ & \textbf{97.77} ${\scriptstyle \pm 0.12}$ & \textbf{98.96} ${\scriptstyle \pm 0.39}$ & \textbf{87.65} ${\scriptstyle \pm 1.88}$ \\
\bottomrule
\end{tabular}
\end{adjustbox}
\end{center}
\end{table}
\vspace{-4mm}
\section{Conclusion}
\label{sec: conclusion}

Our \emph{MusicSlots} model is the first method to extend object-centric learning to the domain of music. To evaluate such models, we introduced novel multi-object music datasets based on Western tonal music. 
\emph{MusicSlots} successfully decomposes chord spectrograms into their constituent note spectrograms, and outperforms several well-established unsupervised and supervised baselines on downstream note property prediction tasks. 
Representations learned by \emph{MusicSlots} are potentially useful for practical applications, such as music transcription/generation and building more human-like perceptual models of audio and music.

\paragraph{Acknowledgments.}
We thank Hamza Keurti and Yassine Taoudi Benchekroun for insightful discussions. This research was funded by Swiss National Science Foundation grant: 200021\_192356, project NEUSYM and the ERC Advanced grant no: 742870, AlgoRNN. We also thank NVIDIA Corporation for donating DGX machines as part of the Pioneers of AI Research Award.

\bibliography{references}
\bibliographystyle{unsrtnat}

\newpage

\appendix

\section{Experimental Details}
\label{sec: exp-details}
In this section, we report details on our multi-object music datasets, model implementation, and experimental setting.   

\subsection{Multi-Object Music Dataset}
\label{sec: multi-object music dataset details}

\paragraph{Bach Chorales.} The original dataset consists of training, validation and test splits with 229, 76 and 77 chorales respectively.
Each chorale is represented as a sequence of four MIDI values for the Bass, Tenor, Alto and Soprano (BTAS) voices.
If a voice is silent at a given time step, the pitch value is 0. 
Since we are interested in extracting unique chords from the chorales, we first concatenate all MIDI sequences of the 376 chorales together across time, and exclude all the columns corresponding to duplicate chords or single note examples, giving us 3131 unique chords in total.
We then randomly shuffle the dataset and partition it into training, validation and test splits, using a train-validation-test split ratio of 70/20/10.
The MIDI pitch values for the chorales are available at \url{https://github.com/czhuang/JSB-Chorales-dataset}.

\paragraph{JazzNet.} The JazzNet dataset (\url{https://github.com/tosiron/jazznet}) contains 5525 annotated chords (including the inversions).
Of the 5525 chords, we use 2227 unique chords with the MIDI pitch values of their notes ranging from 36 (C2) to 96 (C7).
Similar to the chords in the Bach Chorales, the pitch values of the JazzNet chords are represented as an array of 4 MIDI values, with 0 denoting silence. 
The train-validation-test splits are defined such that the training and validation sets contain only dyads (2-notes) and triads (3-notes), while the test set contains only tetrads (4-notes).

In the the following paragraphs we describe the details of the pipeline to generate our multi-object music datasets starting with the MIDI tokens of chords and finally getting chord/note-level spectrograms and binary masks.

\begin{figure}[!htp]
    \centering
\includegraphics[width=\textwidth]{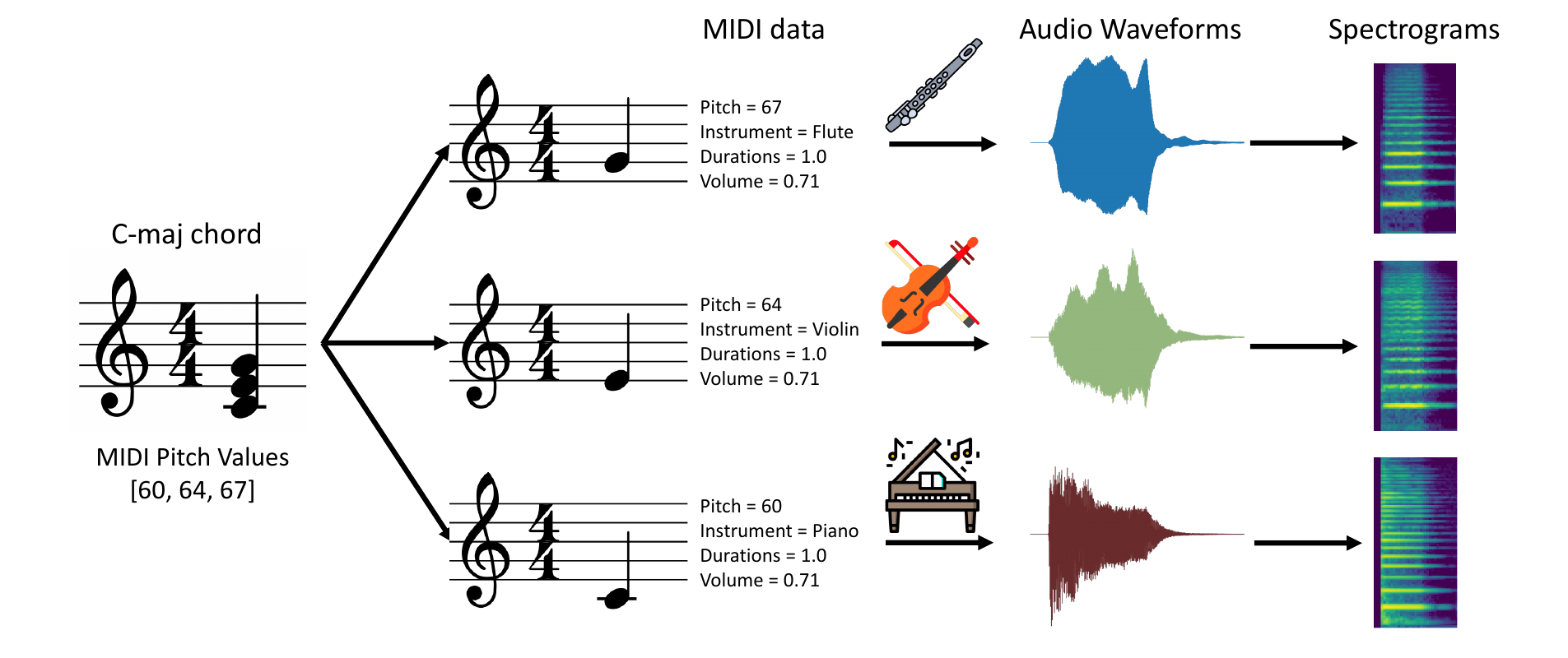}
    \caption{Multi-instrument dataset generation pipeline. Every note in the chord can be synthesized into a raw audio waveform using a different instrument (e.g. Yamaha grand piano, violin, flute).} 
    \label{fig: multi-instrument dataset generation}
\end{figure}

\paragraph{MIDI File Generation.} To allow fine-grained control over the note properties, we generate the MIDI files for the chords ourselves using the Music21 library.
Our data sources define the chords and their pitch values, and we specify the rest of their note attributes (i.e. volume, instrument, duration), as shown in Figure \ref{fig: multi-instrument dataset generation}. 
In the Music21 library, the volume of the note is a scalar value ranging from 0.0 to 1.0 while the duration of the note is measured in seconds. 
We keep the duration and volume fixed across datasets by setting them to 0.71 and 1.0 second respectively.
To generate examples for the multi-instrument version of our dataset, we include the option to change the instrument that plays each note in a chord.  
The list of instruments is defined by a sf2 file.
The sf2 file that is used in our dataset can be downloaded here: \url{https://member.keymusician.com/Member/FluidR3_GM/index.html}.

\paragraph{Audio Waveform Generation.} 
The generated MIDI files are then synthesized into audio waveforms using Fluidsynth (\url{https://www.fluidsynth.org/}). The default PCM quantization settings used in the Fluidsynth library are bit-depth of 16 and sample rate of 44.1 kHz.
We further downsample the waveforms to 16kHz.
We zero-pad the waveforms at the start with a padding size of 4000, which corresponds to about 0.1s of silence.

\paragraph{Waveform to Spectrogram Conversion.} 
We obtain the spectrograms for the chords and their notes by converting their audio waveforms into mel-spectrograms using the TorchAudio library (\url{https://pytorch.org/audio/stable/index.html}). 
We set the number of mel-filter banks to 128 and use the FFT and window sizes of 1024 and hop length of 512. The resulting $128 \times 35$ spectrogram is resized by cropping along the width boundaries [0, 32], giving us a resolution of $128 \times 32$.
To generate a chord spectrogram, we combine the waveforms of its constituent notes, and convert the summed waveform to a mel-spectrogram using the same mel-spectrogram parameters.

\begin{figure}[!htb]
    \centering
    \includegraphics[width=0.75 \textwidth]{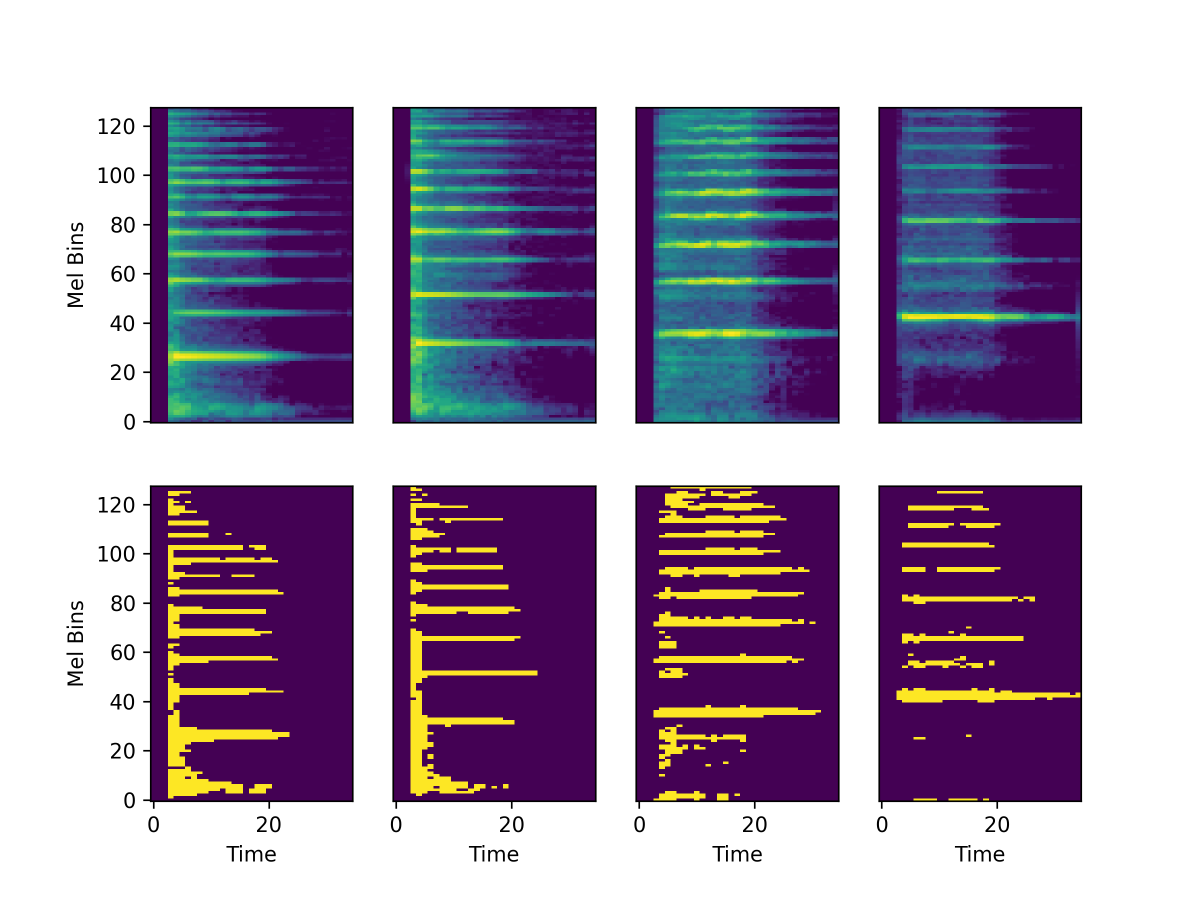}
    \caption{Visualization of the note spectrograms (top row) and the corresponding decibel-thresholded binary masks (bottom row)}
    \label{fig: binary mask visualization}
\end{figure}

\paragraph{Mask Generation.} To generate the binary masks from the mel-spectrograms, we use a fixed decibel threshold value of -30 dB for both the ground-truth and the predicted note spectrograms. Examples of the binary masks for different note spectrograms are shown in Figure \ref{fig: binary mask visualization}. 

\paragraph{Dataset Statistics.}
In Tables \ref{table: summary of chord statistics in Bach Chorales} and \ref{table: summary of chord statistics in JazzNet}, we report the numbers of dyads, triads and tetrads in the datasets. The dataset splits, number of unique pitch values (represented as MIDI note numbers) and instruments are summarized in Table \ref{table: datasets summary}.

\begin{table}[!htp]
\centering
\caption{Chord statistics for Bach Chorales.}
\begin{tabular}{ccccc}
\toprule
Splits         & Dyads & Triads & Tetrads &  \textbf{Total} \\ \midrule
Train          & 10       & 270       & 1910       & 2190              \\
Validation     & 1        & 85        & 540        & 626               \\
Test           & 1        & 43        & 271        & 315               \\ \midrule
\textbf{Total} & 12       & 398       & 2721       & 3131              \\ \bottomrule
\end{tabular}
\label{table: summary of chord statistics in Bach Chorales}
\end{table}

\begin{table}[!htb]
\centering
\caption{Chord statistics for JazzNet}
\begin{tabular}{ccccc}
\toprule
Splits            & Dyads & Triads & Tetrads & \textbf{Total} \\ \midrule
Train             & 530      & 544       & 0          & 1074              \\
Validation        & 124      & 145       & 0          & 269               \\
Test              & 0        & 0         & 884        & 884               \\ \midrule
\textbf{Total} & 654      & 689       & 884        & 2227              \\ \bottomrule
\end{tabular}
\label{table: summary of chord statistics in JazzNet}
\end{table}

\begin{table}[!htb]
\centering
\caption{Number of examples in the dataset splits, number of unique pitch values and name of the instruments used.}
\begin{tabular}{cccccc}
\toprule
Dataset Name   & Train & Validation & Test  & Pitch Values & Instrument(s)        \\ \midrule
JSB-single     & 2190  & 626        & 315   & 53           & Piano                \\
JSB-multi      & 19719 & 5634       & 2826  & 53           & Piano, Violin, Flute \\
Jazznet-single & 1074  & 269        & 884   & 62           & Piano                \\
Jazznet-multi  & 19458 & 5031       & 71604 & 62           & Piano, Violin, Flute \\ \bottomrule
\end{tabular}
\label{table: datasets summary}
\end{table}

\newpage
\subsection{Model Architecture Details}
\label{sec: model-details}
Here we describe the architectural details of all models used in this work.

\paragraph{Convolutional Encoder.} Table \ref{table: MusicSlots Encoder} describes the model architecture for the CNN encoder of the MusicSlots. We use a CNN encoder similar to the one found in \cite{locatello2020slotattn} for the unsupervised object discovery task. 
All convolution layers use a kernel size of $5\times 5$ with a channel size of 128. 
Unlike \cite{locatello2020slotattn}, we set the stride for the horizontal axis to 2.
We find that this improves performance for the unsupervised note discovery task (see Table \ref{table: multi-inst ablation} for details).

\begin{table}[htb!]
\centering
\caption{CNN Encoder in MusicSlots.}
\begin{tabular}{ccccc}
\toprule
Layer              & \begin{tabular}[c]{@{}c@{}}Feature Dimension\\ $H \times W \times C$\end{tabular} & Activation & Stride & \begin{tabular}[c]{@{}c@{}}Padding\\ Input / Output\end{tabular} \\ \midrule
Input              & $128 \times 32 \times 1$                                                          & -          & -      & -                                                                \\ \midrule
Conv $5\times5$    & $128 \times 16 \times 128$                                                        & ReLU       & (1, 2) & (2, 2) / -                                                       \\
Conv $5\times5$    & $128 \times 8 \times 128$                                                         & ReLU       & (1, 2) & (2, 2) / -                                                       \\
Conv $5\times5$    & $128 \times 4 \times 128$                                                         & ReLU       & (1, 2) & (2, 2) / -                                                       \\
Conv $5\times5$    & $128 \times 2 \times 128$                                                         & ReLU       & (1, 2) & (2, 2) / -                                                       \\
Position Embedding & $128 \times 2 \times 128$                                                         & -          & -      & -                                                                \\
Flatten            & $1 \times 256 \times 128$                                                         & -          & -      & -                                                                \\
Layer Norm         & $1 \times 256 \times 128$                                                         & -          & -      & -                                                                \\
Linear             & $1 \times 256 \times 128$                                                         & ReLU       & -      & -                                                                \\
Linear             & $1 \times 256 \times 128$                                                         & -          & -      & -                                                                \\ \bottomrule
\end{tabular}
\label{table: MusicSlots Encoder}
\end{table}

\paragraph{Positional Embedding.} We use the same positional embeddings as \cite{locatello2020slotattn}. 
The positional embedding is a $W \times H \times 4$ tensor, where $W$ and $H$ are width and height of the CNN feature maps respectively. 
The positional information is defined by a linear gradient [0, 1] in each of the four cardinal directions.
Essentially, every point on the grid is a four-dimensional vector that indicates its relative distance to the four edges of the feature map.
We define a learnable linear projection that projects the feature vectors to match the dimensionality of the CNN feature vectors. 
We finally add the linearly projected result to the input CNN feature maps.

\paragraph{Slot Attention Module.} For all experiments, we use the same number of slots $K = 7$ and slot attention iterations $T = 3$.
We set $D$ and $D_{s}$ to be 128 for the dimensions of the linear projections and the slots respectively.
The hidden state of the GRU cell has a dimension of 128.
The residual MLP has a single hidden layer of size 128 with ReLU activation, followed by a linear layer.

\paragraph{De-convolutional Decoder.} We follow the same spatial broadcast deconvolutional decoder (\cite{watters2019spatial}) used in \citep{locatello2020slotattn}, except we set the number of channels in the transposed convolution layers to 128. The overall architecture for the MusicSlots decoder is detailed in Table \ref{table: MusicSlots Decoder}.

\begin{table}[htp!]
\centering
\caption{Deconvolution-based slot decoder in MusicSlots.}
\begin{tabular}{ccccc}
\toprule
Layer                    & \begin{tabular}[c]{@{}c@{}}Feature Dimensions\\ $K \times H \times W \times C$\end{tabular} & Activation & Stride & \begin{tabular}[c]{@{}c@{}}Padding\\ Input / Output\end{tabular} \\ \midrule
Input                    & $7 \times 1 \times 1 \times 128$                                                            & -          & -      & -                                                                \\ \midrule
Spatial Broadcast        & $7 \times 8 \times 2 \times 128$                                                            & -          & -      & -                                                                \\
Position Embedding       & $7 \times 8 \times 2 \times 128$                                                            & -          & -      & -                                                                \\
ConvTranspose $5\times5$ & $7 \times 16 \times 4 \times 128$                                                           & ReLU       & (2, 2) & (2, 2) / (1, 1)                                                  \\
ConvTranspose $5\times5$ & $7 \times 32 \times 8 \times 128$                                                           & ReLU       & (2, 2) & (2, 2) / (1, 1)                                                  \\
ConvTranspose $5\times5$ & $7 \times 64 \times 16 \times 128$                                                          & ReLU       & (2, 2) & (2, 2) / (1, 1)                                                  \\
ConvTranspose $5\times5$ & $7 \times 128 \times 32 \times 128$                                                         & ReLU       & (2, 2) & (2, 2) / (1, 1)                                                  \\
ConvTranspose $5\times5$ & $7 \times 128 \times 32 \times 128$                                                         & ReLU       & (1, 1) & (2, 2) / -                                                       \\
ConvTranspose $3\times3$ & $7 \times 128 \times 32 \times 1$                                                           & -          & (1, 1) & (1, 1) / -                                                                                               \\ \bottomrule
\end{tabular}
\label{table: MusicSlots Decoder}
\end{table}

\paragraph{Baseline AutoEncoders.} The architectural details for the encoder and decoder of the baseline AutoEncoders (AutoEncoder, VAE, $\beta$-VAE) are presented in Tables \ref{table: AE baselines encoder} and \ref{table: AE baselines decoder}. 
We set the latent space dimension for the baseline AutoEncoders to 128. 

\begin{table}[!htp]
\centering
\caption{Convolutional encoder for the baseline AutoEncoders, excluding the final two Linear layers that parameterize the $\mu$ and $\sigma$ of the approximate posterior for the VAE.}
\begin{tabular}{ccccc}
\toprule
Layer                & \begin{tabular}[c]{@{}c@{}}Feature Dimension\\ $H \times W \times C$\end{tabular} & Activation & Stride & \begin{tabular}[c]{@{}c@{}}Padding\\ Input / Output\end{tabular} \\ \midrule
Input                & $128 \times 32 \times 1$                                                          & -          & -      & -                                                                \\ \midrule
Conv $5\times5$      & $64 \times 16 \times 128$                                                         & ReLU       & (2, 2) & (2, 2) / -                                                       \\
Conv $5\times5$      & $32 \times 8 \times 128$                                                          & ReLU       & (2, 2) & (2, 2) / -                                                       \\
Conv $5\times5$      & $16 \times 4 \times 128$                                                          & ReLU       & (2, 2) & (2, 2) / -                                                       \\
Conv $5\times5$      & $8 \times 2 \times 128$                                                           & ReLU       & (2, 2) & (2, 2) / -                                                       \\
Flatten              & $1 \times 1 \times 2048$                                                          & -          & -      & -                                                                \\ \bottomrule
\end{tabular}
\label{table: AE baselines encoder}
\end{table}

\begin{table}[htb!]
\centering
\caption{De-convolutional decoder for the AutoEncoder baselines.}
\begin{tabular}{ccccc}
\toprule
Layer                    & \begin{tabular}[c]{@{}c@{}}Feature Dimensions\\ $H \times W \times C$\end{tabular} & Activation & Stride & \begin{tabular}[c]{@{}c@{}}Padding\\ Input / Output\end{tabular} \\ \midrule
Input                    & $1 \times 1 \times 128$                                                            & -          & -      & -                                                                \\ \midrule
Linear                   & $1 \times 1 \times 2048$                                                           & -          & -      & -                                                                \\
Reshape                  & $8 \times 2 \times 128$                                                            & -          & -      & -                                                                \\
ConvTranspose $5\times5$ & $16 \times 4 \times 128$                                                           & ReLU       & (2, 2) & (2, 2) / (1, 1)                                                  \\
ConvTranspose $5\times5$ & $32 \times 8 \times 128$                                                           & ReLU       & (2, 2) & (2, 2) / (1, 1)                                                  \\
ConvTranspose $5\times5$ & $64 \times 16 \times 128$                                                          & ReLU       & (2, 2) & (2, 2) / (1, 1)                                                  \\
ConvTranspose $5\times5$ & $128 \times 32 \times 128$                                                         & ReLU       & (2, 2) & (2, 2) / (1, 1)                                                  \\
ConvTranspose $5\times5$ & $128 \times 32 \times 128$                                                         & ReLU       & (1, 1) & (2, 2) / -                                                       \\
ConvTranspose $3\times3$ & $128 \times 32 \times 1$                                                           & -          & (1, 1) & (1, 1) / -                                                       \\ \bottomrule
\end{tabular}
\label{table: AE baselines decoder}
\end{table}

\paragraph{Linear Classifier for MusicSlots.} A  linear classifier is trained on every slot that is matched with a note to independently predict its pitch value and instrument identity. 
The linear classifier outputs two vectors $\hat{y}_{\text{inst}} \in N_{\text{inst}}$ and $\hat{y}_{\text{pitch}} \in N_{\text{pitch}}$, where $N_{\text{pitch}}$ is the number of unique pitch values and $N_{\text{inst}}$ is the number of instruments in the dataset. Both $\hat{y}_{\text{inst}}$ and $\hat{y}_{\text{pitch}}$ are normalized using a softmax activation, since pitch values and instrument identities are encoded as one-hot vectors. 
\paragraph{Linear Classifier for Baseline AutoEncoders.} In the baseline AutoEncoders, the representations of individual notes are not readily available. 
Therefore, the input to the linear classifier is a single latent vector. 
The classifier outputs a prediction $\hat{y} \in N_{\text{inst}} \times N_{\text{pitch}}$ for the properties of all the notes in a chord at once. 
In this case, $\hat{y}$ uses sigmoid activation, since the label $y$ is encoded as a multi-hot vector. 

\paragraph{Supervised CNN.} The model architecture for the supervised baseline CNN is depicted in Table \ref{table: supervised CNN baseline}. It follows the same encoder backbone as the one used in MusicSlots. The output of the encoder module is followed by a 2-layer MLP with an output size $N_{\text{inst}} \times N_{\text{pitch}}$. 

\begin{table}[!htb]
\centering
\caption{Supervised CNN for the property prediction task.}
\begin{tabular}{ccccc}
\toprule
Layer           & \begin{tabular}[c]{@{}c@{}}Feature Dimension\\ $H \times W \times C$\end{tabular} & Activation & Stride & \begin{tabular}[c]{@{}c@{}}Padding\\ Input / Output\end{tabular} \\ \midrule
Input           & $128 \times 32 \times 1$                                                          & -          & -      & -                                                                \\ \midrule
Conv $5\times5$ & $64 \times 16 \times 128$                                                         & ReLU       & (2, 2) & (2, 2) / -                                                       \\
Conv $5\times5$ & $32 \times 8 \times 128$                                                          & ReLU       & (2, 2) & (2, 2) / -                                                       \\
Conv $5\times5$ & $16 \times 4 \times 128$                                                          & ReLU       & (2, 2) & (2, 2) / -                                                       \\
Conv $5\times5$ & $8 \times 2 \times 128$                                                           & ReLU       & (2, 2) & (2, 2) / -                                                       \\
Flatten         & $1 \times 1 \times 2048$                                                          & -          & -      & -                                                                \\ 
Linear          & $1 \times 1 \times 128$                                                           & ReLU       & -      & -                                                                \\
Linear          & $1 \times \text{output size}$                       & Sigmoid          & -      & -                                                                \\ \bottomrule
\end{tabular}
\label{table: supervised CNN baseline}
\end{table}

\subsection{Training Details}
\label{sec: training-details}
In this section, we provide an overview of the training details for MusicSlots and its baselines, including their hyperparameter choices and training objectives.
The hyperparameters for training MusicSlots on the unsupervised note discovery task are shown in Table \ref{table: note discovery training}.
All downstream classifiers, including the supervised CNN model, share the common hyperparameters for the note property prediction task, as shown in Table \ref{table: note property training}.
The training hyperparameters for the unsupervised baselines (i.e. AutoEncoder, VAE, $\beta$-VAE) are displayed in Table \ref{table: unsupervised ae training}. \\
\paragraph{Baseline AutoEncoders.} 
The baseline AutoEncoder follows the same training objective as MusicSlots: they are both trained to minimize the Mean Square Error (MSE) between the predicted and input chord spectrogram $\mathcal{L} = ||\mathbf{x} - \mathbf{\hat{x}}||_2^2$.
The baseline VAE and $\beta$-VAE models are trained by maximizing the evidence lower bound (ELBO), where the weight of the KL-divergence term $\beta$ is set to 1 for the baseline VAE.
For more details on the effect of different choices for $\beta$ on the downstream note property prediction task, please refer to Table \ref{table: effect of beta value in VAE} in \Cref{sec: add-results}.

\paragraph{Downstream Classifiers.} 
The downstream note property classifier for MusicSlots is trained by minimizing the categorical cross-entropy loss. 
For the supervised CNN and the linear classifier trained on the baseline AutoEncoders, we use binary cross-entropy loss, since the target is a multi-hot vector. 

\begin{table}[!htp]
\centering
\caption{Training hyperparameters of the MusicSlots model for unsupervised note discovery experiments}
\begin{tabular}{ll}
\toprule
Hyperparameters              &       \\ \midrule
Training Steps                        & 100K  \\
Batch Size                            & 32 
\\
Optimizer                             & Adam  \\
Max. Learning Rate                    & 1e-04 \\
Learning Rate Warmup Steps            & 10K   \\
Decay Steps                           & 500K  \\
Gradient Norm Clipping                & 1.0   \\ \bottomrule
\label{table: note discovery training}
\end{tabular}
\end{table} 

\begin{table}[!htp]
\centering
\caption{Training hyperparameters for the note property prediction task}
\begin{tabular}{ll}
\toprule
Hyperparameters &       \\ \midrule
Training steps  & 10K   \\
Batch Size      & 32    \\
Optimizer       & Adam  \\
Learning Rate   & 1e-03 \\ \bottomrule
\label{table: note property training}
\end{tabular}
\end{table}

\begin{table}[!htp]
\centering
\caption{Training hyperparameters for the baseline AEs during the unsupervised pre-training.}
\begin{tabular}{ll}
\toprule
Hyperparameters        &       \\ \midrule
Training Steps         & 100K  \\
Learning Rate          & 1e-04 \\
Batch Size             & 32    \\
Optimizer              & Adam  \\
Decay Steps            & 100K  \\
Gradient Norm Clipping & 1.0   \\ \bottomrule
\end{tabular}
\label{table: unsupervised ae training}
\end{table}

\subsection{Evaluation Details}
\label{sec: eval-metrics}
\paragraph{Note Discovery.}
To compute the note MSE, we first calculate the mean squared error between all pairs of predicted and ground-truth note spectrograms. Since the orders of the predictions and the  ground-truth are arbitrary, we match them using the Hungarian algorithm (\cite{Kuhn1955Hungarian}) to find the matching with the lowest MSE. 
mIoU is calculated by first computing all pairwise IoUs between the predicted and
ground-truth dB-thresholded masks, and using the Hungarian algorithm to find the optimal assignment that gives the highest mIoU. 
For the Hungarian matching algorithm, we use the scipy implementation \texttt{scipy.optimize.linear\_sum\_assignment}.

\paragraph{Note Property Prediction.} The performance of the classifiers is quantified using classification accuracy. The accuracy is measured by computing the percentage of correctly classified chord examples in the dataset. 
A chord is considered to be correctly classified if and only if the classifier predictions for all of its note properties (i.e. note pitch values, instrument identities) are correct. 

\section{Additional Results}
\label{sec: add-results}
In this section, we present additional results that quantify the importance of different modelling choices.

\paragraph{Single-instrument Note Discovery Results} Table \ref{table: single instrument note discovery results} shows the unsupervised note discovery performance of our MusicSlots model with different alpha mask normalization choices on the single-instrument JSB and JazzNet datasets. 
We observe significant performance gain in the single-instrument setting when sigmoid-normalized alpha masks or no alpha masks are used in our MusicSlots model. 

\begin{table}[h]
\centering
\caption{Note discovery results on single-instrument BachChorales (JSB) and JazzNet datasets for MusicSlots models with different choices for $f_{norm}$ function. Mean and std-dev. are reported across 5 seeds.}
\begin{adjustbox}{width=0.7\textwidth}
\begin{tabular}{c|c|c|c}
\toprule
 Datasets & Mask Norm. & Note MSE $\downarrow$ & mIoU $\uparrow$ \\ \midrule
\multirow{3}{*}{JSB-single} & MusicSlots-soft & 75.32 ${\scriptstyle \pm 37.63}$ & 0.68 ${\scriptstyle \pm 0.08}$ \\
 & MusicSlots-sigm & \textbf{18.21} ${\scriptstyle \pm 3.40}$ & \textbf{0.83} ${\scriptstyle \pm 0.02}$ \\ & MusicSlots-none & 22.44 ${\scriptstyle \pm 7.07}$ & 0.81 ${\scriptstyle \pm 0.03}$ \\ \midrule
 \multirow{3}{*}{JazzNet-single} & MusicSlots-soft & 114.09 ${\scriptstyle \pm 22.72}$ & 0.63 ${\scriptstyle \pm 0.04}$ \\
 & MusicSlots-sigm & 49.05 ${\scriptstyle \pm 5.98}$ & 0.75 ${\scriptstyle \pm 0.03}$ \\ & MusicSlots-none & \textbf{44.49} ${\scriptstyle \pm 0.62}$ & \textbf{0.76} ${\scriptstyle \pm 0.02}$ \\
\bottomrule
\end{tabular}
\end{adjustbox}
\newline
\label{table: single instrument note discovery results}
\end{table}

\paragraph{Ablation on Architectural Modifications} Table \ref{table: multi-inst ablation} presents our ablation study on different architectural choices in our MusicSlots model on the multi-instrument Bach Chorales and JazzNet datasets. 
We start from the `Default' model that follows the same setup used for object discovery in \cite{locatello2020slotattn}. 
In this setup, we have stride length of (1, 1) in the convolutional encoder layers and the alpha masks of the decoder are normalized using the \texttt{Softmax} function.
We find that both implicit differentiation (\cite{chang2022fixedpoints}) and the removal of the alpha masks from the decoder play a crucial role in improving the note discovery performance of our MusicSlots model.
Increasing the stride length along the time axis to 2 also improves its performance, though not as significantly as the other two design choices. 
By combining these improvements in the model architecture and training optimization, we finally arrive at our MusicSlots model without any alpha masking.

\begin{table}[!htb]
\centering
\caption{Architectural ablations on the MusicSlots for unsupervisd note discovery task. `Default' here refers to the MusicSlots model with the setup used for object discovery in \cite{locatello2020slotattn} , where stride = (1, 1) in the convolutional encoder layers and the spatial broadcast decoder outputs softmax alpha masks.}
\begin{tabular}{cccc}
\toprule
Dataset                        & Model                              & Note MSE $\downarrow$ & mIoU $\uparrow$ \\ \midrule
\multirow{5}{*}{JazzNet-multi}       & Default                            & 70.05 $\scriptstyle {\pm 23.61}$     & 0.70 $\scriptstyle {\pm 0.07}$ \\
                               & Default + stride\_length = (1, 2)  & 51.31 $\scriptstyle {\pm 1.89}$      & 0.76 $\scriptstyle {\pm 0.04}$ \\
                               & Default - Softmax Alpha Mask       & 39.08 $\scriptstyle {\pm 6.35}$      & 0.79 $\scriptstyle {\pm 0.02}$ \\
                               & Default + Implicit Differentiation & 32.56 $\scriptstyle {\pm 9.84}$      & 0.83 $\scriptstyle {\pm 0.00}$ \\
                               & MusicSlots                         & \textbf{19.95} $\scriptstyle {\pm 1.89}$      & \textbf{0.90} $\scriptstyle {\pm 0.01}$ \\ \midrule
\multirow{5}{*}{JSB-multi} & Default                            & 100.77 $\scriptstyle {\pm 52.91}$    & 0.60 $\scriptstyle {\pm 0.15}$ \\
                               & Default + stride\_length = (1, 2)  & 60.22 $\scriptstyle {\pm 14.56}$     & 0.76 $\scriptstyle {\pm 0.04}$ \\
                               & Default - Softmax Alpha Mask       & 40.06 $\scriptstyle {\pm 14.60}$     & 0.78 $\scriptstyle {\pm 0.04}$ \\
                               & Default + Implicit Differentiation & 59.34 $\scriptstyle {\pm 22.01}$     & 0.79 $\scriptstyle {\pm 0.04}$ \\
                               & MusicSlots                         & \textbf{13.47} $\scriptstyle {\pm 0.90}$      & \textbf{0.91} $\scriptstyle {\pm 0.01}$ \\ \bottomrule
\end{tabular}
\label{table: multi-inst ablation}
\end{table}

\paragraph{VAE Ablation} Table \ref{table: effect of beta value in VAE} shows the ablation study for the choice of $\beta$ in the VAEs on the multi-instrument Bach Chorales and JazzNet datasets.
We observe that higher $\beta$ results in worse downstream property prediction performance on both datasets and the best results are achieved using $\beta = 0.5$. 

\begin{table}[!htb]
\centering
\caption{Note property prediction performance of baseline VAEs with different $\beta$ values. $\beta = 1$ corresponds to the vanilla VAE model.
As we increase $\beta$, the property prediction performance using the latent representations from the $\beta$-VAE worsens on both JSB and Jazznet multi-instrument datasets.}
\begin{tabular}{c|cc|cc}
\toprule
\multirow{2}{*}{$\beta$} & \multicolumn{2}{c|}{JSB-multi}                                                      & \multicolumn{2}{c}{Jazznet-multi}                                                   \\ \cline{2-5} 
                         & Val-Acc.                                 & Test-Acc.                                & Val-Acc.                                 & Test-Acc.                                \\ \midrule
0.5                      & \textbf{97.85} ${\scriptstyle \pm 0.13}$ & \textbf{97.42} ${\scriptstyle \pm 0.06}$ & \textbf{97.00} ${\scriptstyle \pm 0.38}$ & \textbf{81.53} ${\scriptstyle \pm 0.77}$ \\
1.0                      & 96.66 ${\scriptstyle \pm 0.34}$          & 96.21 ${\scriptstyle \pm 0.35}$          & 94.37 ${\scriptstyle \pm 0.55}$          & 71.55 ${\scriptstyle \pm 4.77}$          \\
2.0                      & 92.23 ${\scriptstyle \pm 0.66}$          & 90.74 ${\scriptstyle \pm 0.94}$          & 73.93 ${\scriptstyle \pm 1.97}$          & 47.90 ${\scriptstyle \pm 8.59}$          \\
4.0                      & 82.07 ${\scriptstyle \pm 0.83}$          & 78.95 ${\scriptstyle \pm 1.42}$          & 47.51 ${\scriptstyle \pm 3.71}$          & 11.31 ${\scriptstyle \pm 1.92}$          \\ \bottomrule
\end{tabular}
\label{table: effect of beta value in VAE}
\end{table}

\clearpage

\section{Note Discovery Visualization}
\label{sec: add-viz}

\begin{figure}[!htb]
    \centering
    \includegraphics[width=0.8\textwidth]{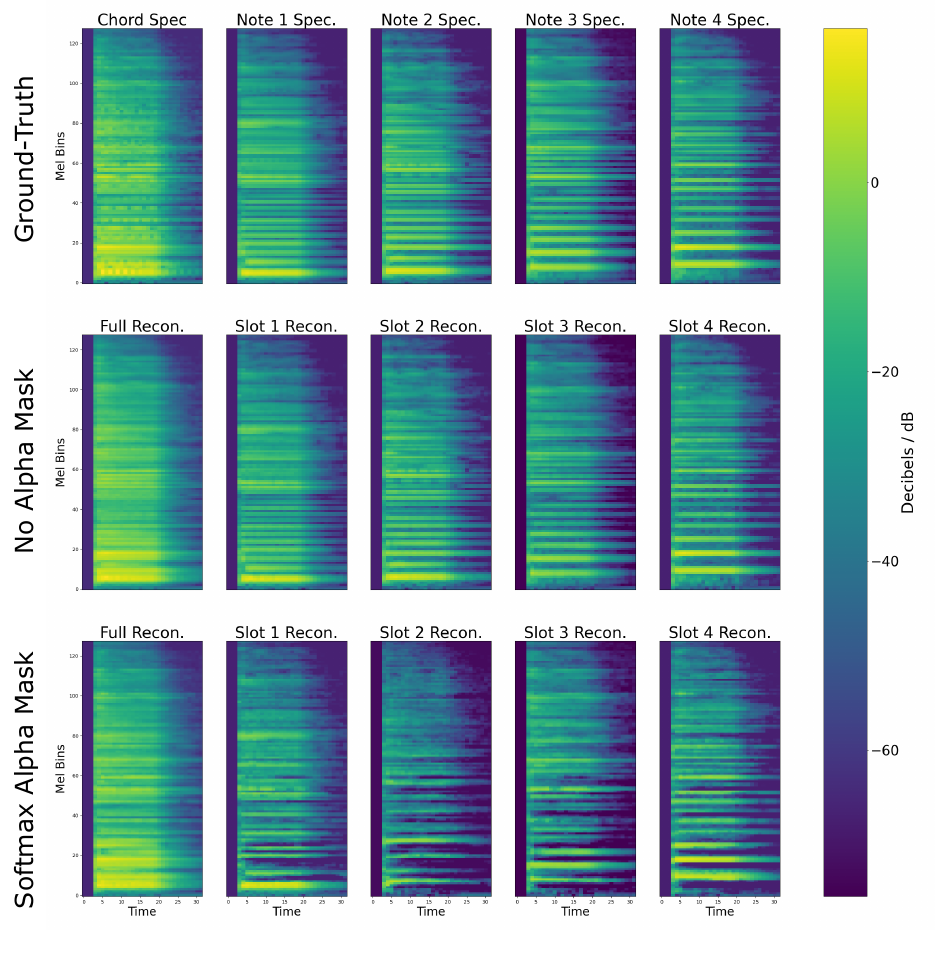}
    \caption{Qualitative performance comparison of MusicSlots without alpha masks (row 2) and MusicSlots with softmax-normalized alpha masks (row 3) on the JazzNet single-instrument dataset. MusicSlots with softmax alpha masks leaves unnatural gaps in the lower frequency bins where there is a strong degree of overlap between the note spectrograms. MusicSlots without any alpha mask does not introduce these artifacts and models the overlapping regions more accurately than its softmax normalized counterpart.}
    \label{fig: alpha mask qualitative result}
\end{figure}

We provide additional visualization samples of note discovery results from our MusicSlots model on the JazzNet and Bach Chorales datasets, including both the success and failure cases.
We also visualize the effect of using \texttt{Softmax} alpha masks in MusicSlots for note discovery in Figure \ref{fig: alpha mask qualitative result}.

\begin{figure}[!htb]
    \centering
    \includegraphics[width=0.75\textwidth]{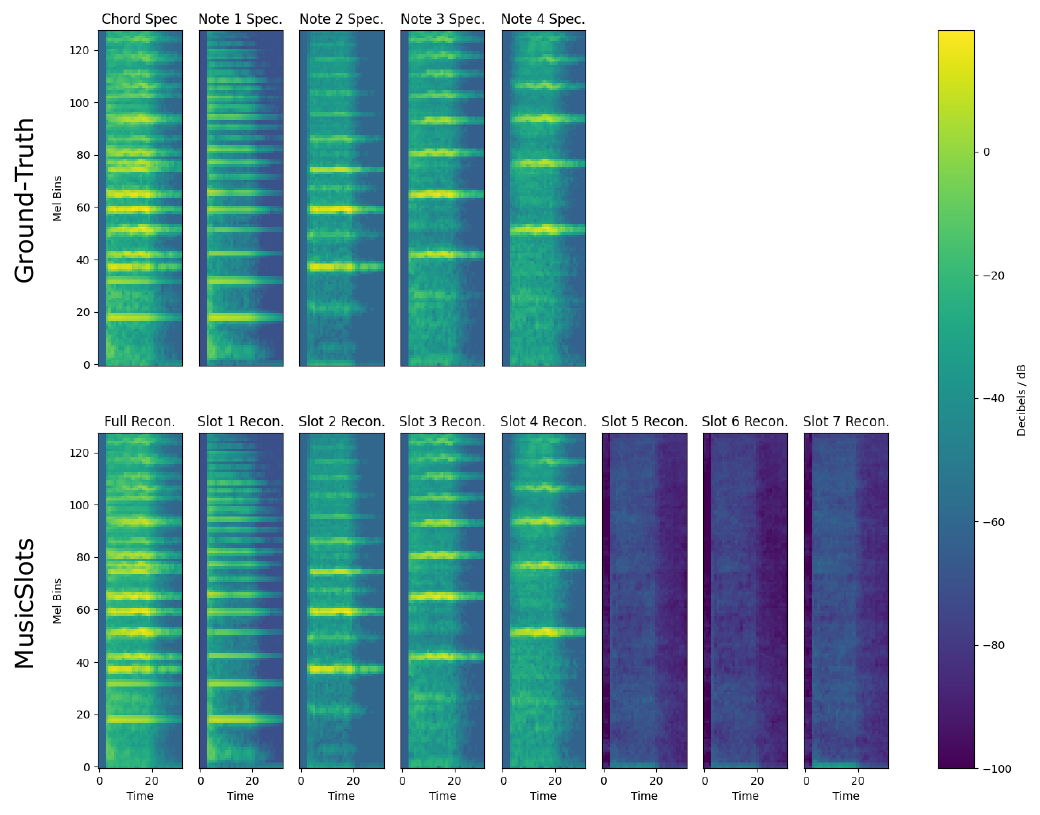}
    \caption{Unsupervised note discovery result on the JSB-multi-instrument dataset. The MusicSlots accurately predicts the ground-truth note spectrograms. It also learns to capture the background (i.e. silence) in the remaining slots that are not matched with the ground-truth notes.}
    \label{fig: jsb_good_result}
\end{figure}

\begin{figure}[!htb]
    \centering    \includegraphics[width=0.75\textwidth]{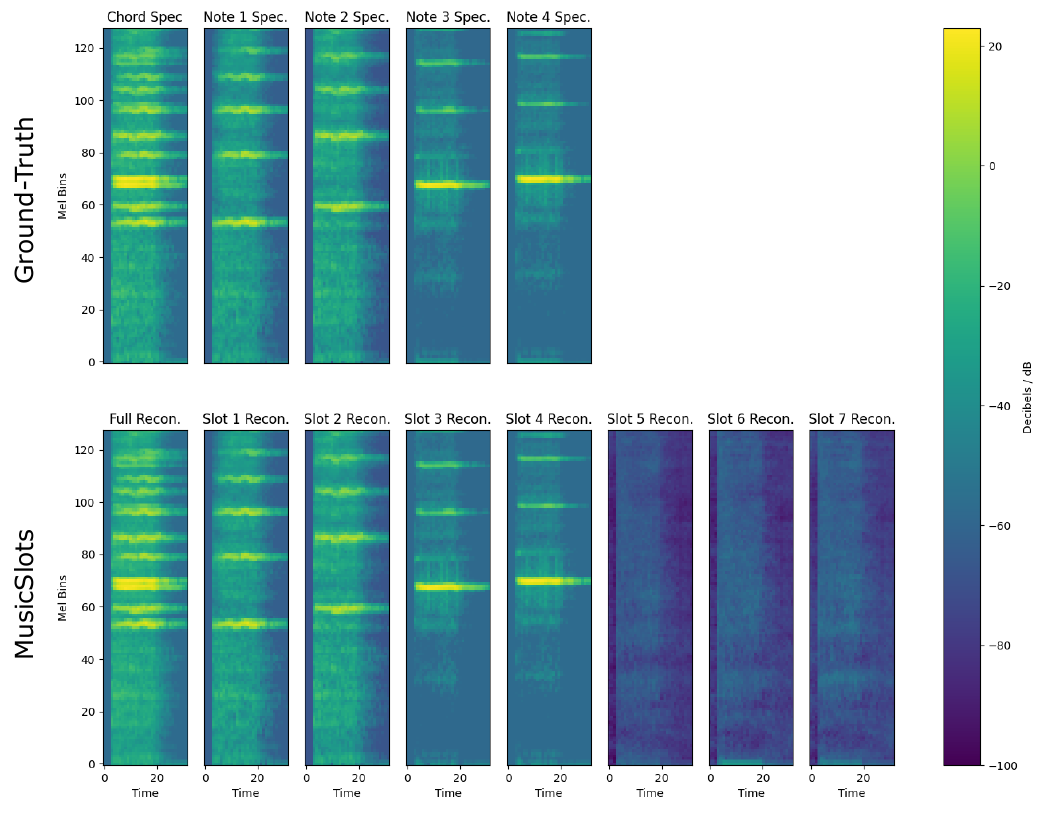}
    \caption{Unsupervised note discovery result on the Jazznet-multi-instrument dataset. Similar to the example visualized in Figure \ref{fig: jsb_good_result}, MusicSlots successfuly decomposes the given chord spectrogram into its constituent note spectrograms and distribute the background across the remaining slots.}
    \label{fig: jazznet_good_result}
\end{figure}

\begin{figure}
    \centering
    \includegraphics[width=0.75\textwidth]{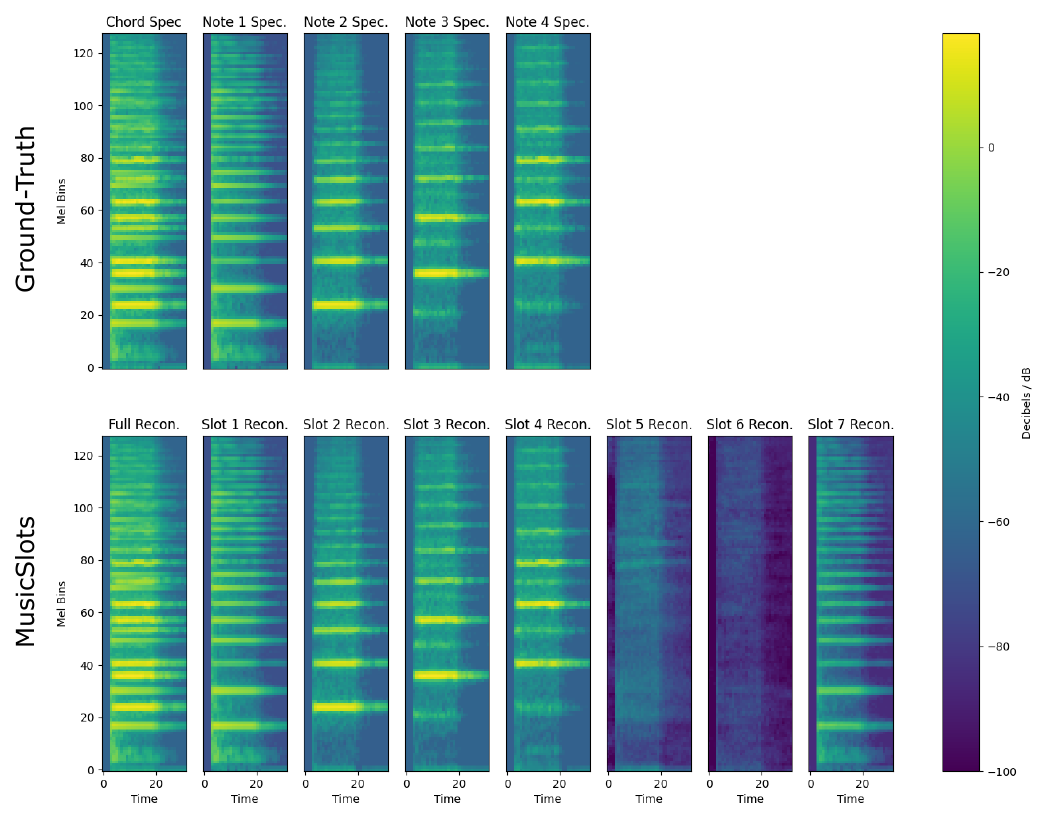}
    \caption{Unsupervised note discovery result on the JSB-multi-instrument dataset. On this example, MusicSlots successfully predicts most of the ground-truth note spectrograms.
    However, it oversegments one of the notes (Note 1) by assigning it to slot 1 and 7.}
    \label{fig: jsb_failure_case_1}
\end{figure}

\begin{figure}[!htb]
    \centering
    \includegraphics[width=0.75\textwidth]{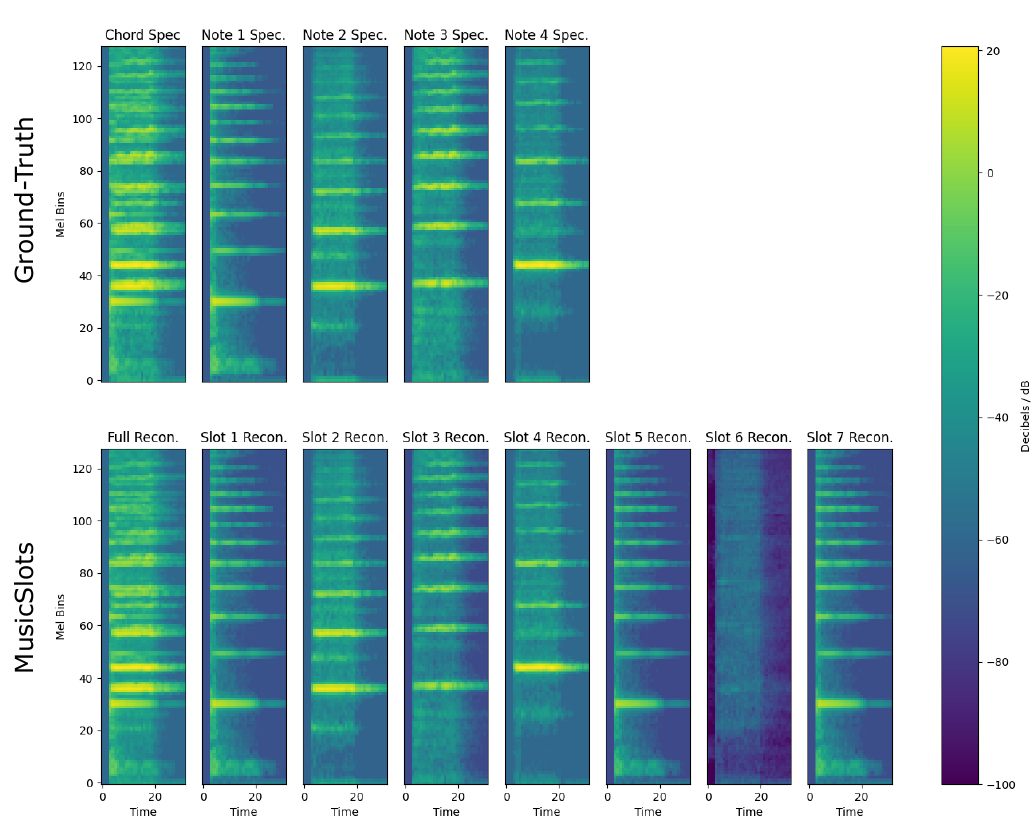}
    \caption{Unsupervised note discovery result on the JazzNet-multi-instrument dataset.
    Similar to the example shown in Figure \ref{fig: jsb_failure_case_1}, MusicSlots performs oversegmentation by using three slots (slot 1, 5 and 7) to model note 1.}
    \label{fig: jazznet_failure_case_1}
\end{figure}

\begin{figure}[!htb]
    \centering
    \includegraphics[width=0.75\textwidth]{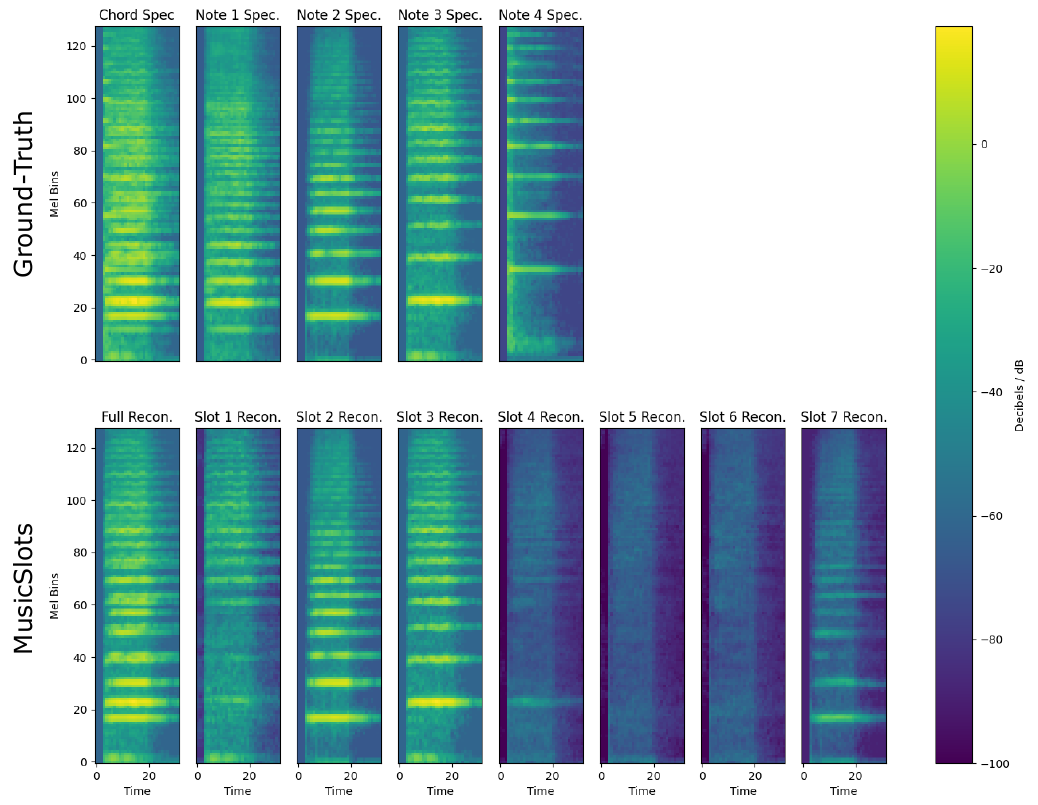}
    \caption{Visualization of a failure case of MusicSlots on the JSB-multi-instrument dataset. Only two of the four matched predictions accurately capture the harmonic structure of the ground-truth note spectrograms.}
    \label{fig: jsb_failure_case_2}
\end{figure}

\begin{figure}[!htb]
    \centering
    \includegraphics[width=0.75\textwidth]{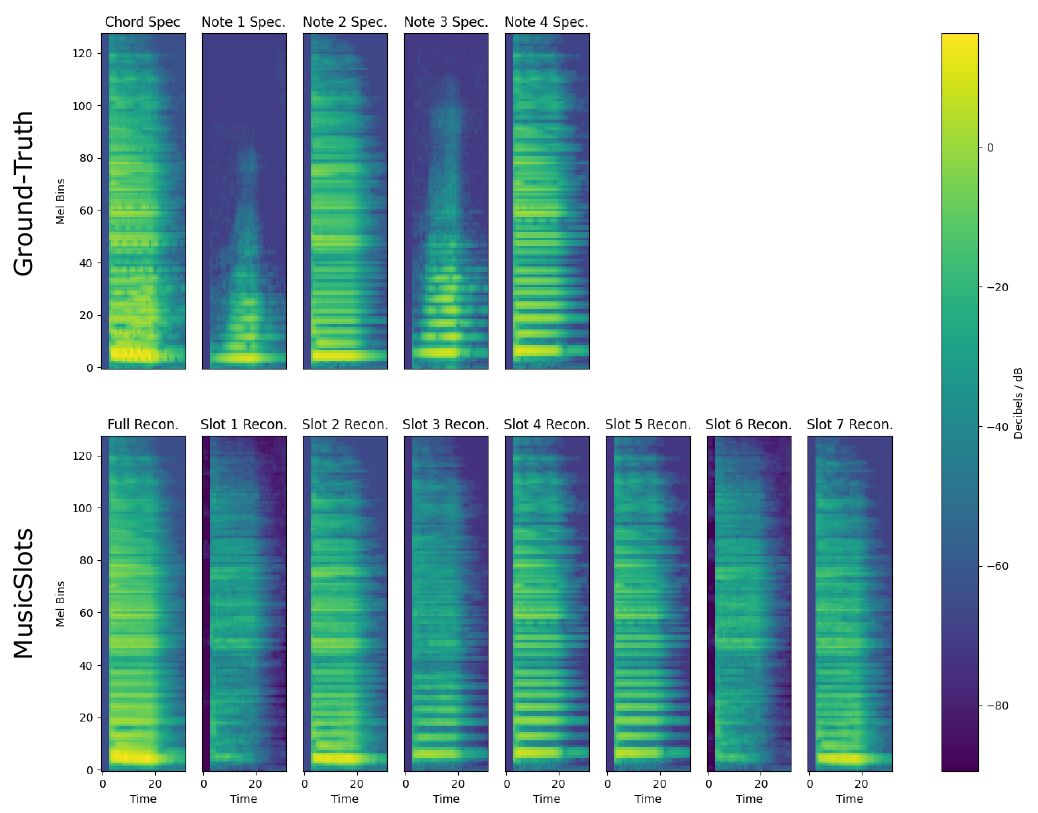}
    \caption{Visualization of a failure case of MusicSlots on the Jazznet-multi-instrument dataset.
    MusicSlots completely fails to predict notes 1 and 3 in its slot reconstructions. 
    It also fails to model the background in any of the slots.}
    \label{fig: jazznet_failure_case_2}
\end{figure}

\end{document}